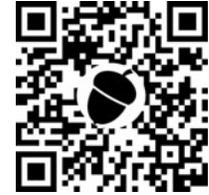

Open camera or QR reader and
scan code to access this article
and other resources online.

# Venus' Atmospheric Chemistry and Cloud Characteristics Are Compatible with Venusian Life

William Bains,[1,2] Janusz J. Petkowski,[1,3] and Sara Seager[1,4,5]

## Abstract

Venus is Earth's sister planet, with similar mass and density but an uninhabitably hot surface, an atmosphere with a water activity 50–100 times lower than anywhere on Earths' surface, and clouds believed to be made of concentrated sulfuric acid. These features have been taken to imply that the chances of finding life on Venus are vanishingly small, with several authors describing Venus' clouds as "uninhabitable," and that apparent signs of life there must therefore be abiotic, or artefactual. In this article, we argue that although many features of Venus can rule out the possibility that Earth life could live there, none rule out the possibility of all life based on what we know of the physical principle of life on Earth. Specifically, there is abundant energy, the energy requirements for retaining water and capturing hydrogen atoms to build biomass are not excessive, defenses against sulfuric acid are conceivable and have terrestrial precedent, and the speculative possibility that life uses concentrated sulfuric acid as a solvent instead of water remains. Metals are likely to be available in limited supply, and the radiation environment is benign. The clouds can support a biomass that could readily be detectable by future astrobiology-focused space missions from its impact on the atmosphere. Although we consider the prospects for finding life on Venus to be speculative, they are not absent. The scientific reward from finding life in such an un-Earthlike environment justifies considering how observations and missions should be designed to be capable of detecting life if it is there. Key Words: Venus—Life—Astrobiology—Habitability—Acidity—Aridity. Astrobiology 23, xxx–xxx.

## 1. Introduction

**T**HE TENTATIVE DETECTION of phosphine in the atmosphere of Venus (Bains *et al.*, 2021b, 2022a; Greaves *et al.*, 2021a, 2021b, 2021c, 2022; Mogul *et al.*, 2021b) has re-ignited interest in the idea that life might exist in Venus' clouds. With that renewed interest have come a variety of suggestions that what we know about Venus inherently rules out the possibility of life, and consequently that investing effort in modeling or detecting Venusian life is a waste of resources at best, unscientific at worst (Cockell *et al.*, 2021b).

In this article, we address those arguments, and show that while life on Venus remains speculative, and although the majority of the community believe that there is only a small chance of that there is life in the clouds of Venus (Bains and Petkowski, 2021), none of the arguments rule out the possibility of life there. We will argue that what we know about Venus does not render the hypothesis that there might be life there unworthy of pursuit.

It is widely assumed that, if present, life can only exist in Venus' clouds, and not on Venus' surface (*e.g.*, Limaye *et al.*, 2018, 2021; Kotsyurbenko *et al.*, 2021; Mogul *et al.*, 2021a; Seager *et al.*, 2021b). Venus' surface is too hot for many organic chemicals to be stable, and there is no naturally occurring substance that would be liquid under Venus' surface conditions to provide a solvent for life. By contrast, Venus' clouds have a temperature range of −20°C to 100°C. At these temperatures, complex chemistry is potentially stable and some components of Venus' atmosphere could be liquid. The cloud layer is therefore sometimes called ''temperate,'' although this statement is misleading, because the conditions in the cloud layers are staggeringly unlike any inhabited terrestrial environment (Bains *et al.*, 2021a, 2021b; Seager *et al.*, 2021b).

[1]Department of Earth, Atmospheric and Planetary Sciences, Massachusetts Institute of Technology, Cambridge, Massachusetts, USA.
[2]School of Physics and Astronomy, Cardiff University, Cardiff, United Kingdom.
[3]JJ Scientific, Warsaw, Poland.
Departments of [4]Physics and [5]Aeronautics and Astronautics, Massachusetts Institute of Technology, Cambridge, Massachusetts, USA.

Several authors have said that the conditions on Venus are so unlike those known to support life on Earth that the probability that there is life on Venus is vanishingly small, and as a consequence they explicitly state that Venus is ''uninhabitable'' (Kane *et al.*, 2019; Cockell *et al.*, 2021a; Hallsworth *et al.*, 2021). In this article, we argue that the fact that the clouds of Venus are so different from the inhabited Earth does not reliably inform our estimate of whether they are habitable. The argument that the clouds are uninhabitable because terrestrial life could not inhabit them relies on the assumptions that all terrestrial life is known, and that all life everywhere must be chemically similar to terrestrial life. Neither assumption is proven, and the second assumption in particular is problematic, extrapolating a single example of life to all possible life. Rather, terrestrial life should be a guide to basic principles to which Venusian life is assumed to adhere, because there is no other model of life available on which to base assumptions, and then explore whether the Venusian clouds are ruled out as a habitat by those principles (Duzdevich *et al.*, 2022).

We will not discuss the controversial claims that global features of Venusian atmosphere are indicators that there is life in the clouds (*e.g.*, Schulze-Makuch *et al.*, 2004; Schulze-Makuch and Irwin, 2006; Limaye *et al.*, 2018; Skladnev *et al.*, 2021). We are solely concerned with whether features of Venus' atmosphere and clouds make it vanishingly improbable for life to exist there. To this end, we discuss how much biomass a Venusian cloud biosphere could comprise, and whether that biomass could produce a detectable perturbation in the atmosphere.

From the biomass estimates, we discuss potential limitations on that biomass, including energy requirements, the low water activity in the atmosphere, the challenges that a low hydrogen environment pose to the population of the chemical space of biochemistry, the acidity of the cloud droplets, the potential lack of metals, and the high radiation environment. We find that none of these is an insuperable conceptual barrier to life in the clouds, although most rule out the possibility of Earth-like life in the clouds of Venus. We end with a discussion about why the hypothesis that life might exist in the clouds of Venus is worthwhile considering, and justifies the direct exploration of the clouds with *in situ* probes.

## 2. The Cloud Habitat of Venus

Terrestrial life is overwhelmingly found on Earth's surface or in bulk liquid water. The clouds of Venus provide a completely different environment, with only microscopic volumes of liquid and no solid surface. Earth supports a large biomass because the planet literally supports it, on land or in liquid water and even in the subsurface. The physical surface of Venus is likely to be uninhabitable. Life requires complex chemistry and a liquid solvent (Hoehler *et al.*, 2020); the surface of Venus is too hot for most complex covalent chemistry to be stable, and no naturally occurring liquid solvent could be stable under surface conditions. The habitable region of Venus is, therefore, the ''temperate'' cloud decks, and specifically the cloud particles. In this section, we discuss why the cloud habitat is not a barrier to the existence of a biosphere that could have a material, detectable effect on the atmosphere.

### 2.1. Potential mass of a cloud-based biosphere

Models of life in the clouds assume that a fraction of the volume of a subset of cloud particles are occupied by micron-sized organisms that can be thought of as analogous in size to terrestrial bacteria. Three model scenarios are summarized in Table 1. We use the cloud particle distributions derived from the Pioneer Venus Sounder probe data (Knollenberg and Hunten, 1980) as representative of the likely particle distribution in the clouds.

We assume that life must live inside cloud particles, as it is dependent on a liquid environment. A biological particle that is free-floating in the gas phase in the cloud layer is likely to either lose liquid (and so desiccate and be unable to grow) or gain liquid (*i.e.*, become the condensation nucleus of a cloud particle), as discussed in (Seager *et al.*, 2021b). Thus, the correct description of the candidate habitat for life on Venus is not the cloud layer (which has a volume of $\sim 8.7 \cdot 10^{10}$ km$^3$), but the cloud particles (which occupy a relatively smaller volume of $3.9 \cdot 10^2$ km$^3$, or about 2% of the volume of Lake Baikal). Table 1 shows that a cloud-based biosphere on Venus must be substantially smaller than the surface-based biosphere on Earth.

Although it is not an *a priori* requirement that all life makes gaseous products, some of which may be valuable as biosignatures (*e.g.*, Seager *et al.*, 2012; Catling *et al.*, 2018), almost all life on Earth is observed to do so. If Venusian life does make a gaseous product, we can test whether that gas could be made at a rate comparable to terrestrial biosignature gases such as methane, isoprene, or even oxygen. We discuss some specific examples of metabolic processes that change the composition of the atmosphere below (Section 3.1.3.).

As a case study, we took the production of ammonia, which has been hypothesized to be made by potential Venusian microorganisms to neutralize their acidic environment (Bains *et al.*, 2021b), and took as an exemplar organism cyanobacteria, in which the rate of ammonia production has been measured as $4 \cdot 10^{-7}$ g NH$_3$/gram wet weight biomass/second (Burris and Roberts, 1993; Sprőber *et al.*, 2003; Reed *et al.*, 2011). This production rate has been demonstrated to be compatible with the tentative detections of ammonia in Venus' atmosphere (Bains *et al.*, 2021a). The production rate predicted for highest biomass loadings in Table 1 is 10% of the terrestrial production of oxygen ($5 \cdot 10^5$ Tg/year) (Badgley *et al.*, 2019), with more realistic lower biomass models in Table 1 having maximal gas production rates comparable to terrestrial production of methane (200 Tg/year from non-anthropogenic sources) (Kirschke *et al.*, 2013) or isoprene (500 Tg/year) (Zhan *et al.*, 2021).

Thus, even a biomass comprising 0.1% of the total cloud mass could produce substantial amounts of gas, assuming that biomass' primary metabolism was one that generated gaseous products. If the half-life of the gas in Venus' atmosphere was sufficiently long, this would accumulate to detectable levels in the atmosphere; for example, terrestrial

**VENUS IS COMPATIBLE WITH VENUSIAN LIFE** 3Table 1. Potential Mass of a Venusian Cloud Biosphere, Under Different Assumptions[a]

| Assumptions | Mass (mg. cm$^{-2}$) | Total mass in the cloud deck (Tg) | Fraction of total cloud deck mass | Mass as fraction of the mass of the terrestrial biosphere | Maximum flux of gas (Tg/Earth year) | Maintenance energy for entire biosphere (J/sec) |
|---|---|---|---|---|---|---|
| All cloud particles | $1.2 \cdot 10^{-2}$ | $5.5 \cdot 10^4$ | 100% | N/A | N/A | N/A |
| All the Mode 2 particles are life (comparable size to Earth's bacterial cells) are biomass | $8.1 \cdot 10^{-4}$ | $3.7 \cdot 10^3$ | 6.8% | 0.2% | $4.7 \cdot 10^4$ | $9.3 \cdot 10^{11}$ |
| 1.5% of the mass of Mode 3 particles (the largest particles) are biomass (as modeled by Bains et al. [2021a]) | $1.7 \cdot 10^{-4}$ | $7.6 \cdot 10^2$ | 1.4% | 0.04% | $9.6 \cdot 10^3$ | $6.0 \cdot 10^{11}$ |
| 0.1% of the cloud mass is living cells | $1.2 \cdot 10^{-5}$ | $5.5 \cdot 10^1$ | 0.1% | 0.0029% | $6.9 \cdot 10^2$ | $4.1 \cdot 10^{10}$ |

[a]The mass of the clouds was calculated from particle distribution in (Knollenberg and Hunten, 1980), as per (Seager et al., 2021b), assuming a droplet density of 1.4 g cm$^{-3}$. The mass of the Earth's biosphere is assumed to be $1.891 \cdot 10^3$ Pg (Begon et al., 1990). Calculations are for the cloud layers only, with the base of the clouds taken as the lowest altitude with significant Mode 3 particles, 47.3 km. Details of the calculations presented in this table are available in the Supplementary Data S1.

N/A, not applicable.

methane accumulates to remotely detectable concentrations in Earth's atmosphere (Sagan et al., 1993), but isoprene does not because of its rapid destruction by tropospheric photochemistry (Zhan et al., 2021). The specifics of the gas's half-life will depend on the specifics of the gas's chemistry in the Venusian atmosphere, which is beyond this article to explore in detail. Here, we just point out that the flux of gas from a metabolism could match the flux of known biosignature gases on Earth, despite the much smaller overall mass of any Venusian biosphere. Whether accumulated gases constitute a biosignature would depend on whether there was an abiotic source of the gas (e.g., Seager et al., 2012; Catling et al., 2018).

We conclude that Venus' aerial biosphere must be much smaller than the Earth's. However, even such scarce, strictly aerial life could leave a detectable mark on the chemistry of the atmosphere in the clouds.

### 2.2. Maintaining life aloft

One rarely discussed barrier to an exclusively cloud-based habitat is that cloud particles will tend to settle under gravity. Even if there is strong convective movement in the atmosphere, the ultimate fate of the bulk of the cloud particles must be to settle to lower, and hence uninhabitably hot, regions of the atmosphere, as any bulk upward motion of the atmosphere must be balanced by an equal bulk downward motion.

Two mechanisms have been suggested to lift some biological particles back into the clouds, allowing a stable population. The first mechanism is lofting via gravity waves (Seager et al., 2021b). The model described by Seager et al. (2021b) suggests that cloud particles containing organisms will settle to increasingly hot regions of the cloud layer, where the organisms will produce spores. As the cloud particles evaporate or shatter (Bains et al., 2021a) the spores are released into the haze layer below the clouds where they may remain dormant. Modeling based on eddy diffusion rates derived from radio occultation measurements of Venus' atmosphere suggests that a fraction of the spores will then be brought back into the cloud layer by mixing caused by gravity waves in the atmosphere on a time scale of ∼1 (terrestrial) year (Seager et al., 2021b), where they act as cloud condensation nuclei for new droplet formation. (We note that spores of mesophilic terrestrial organisms can remain viable at 100°C under dry conditions for months (Nicholson et al., 2000), so the survival of an organism specifically adapted to this environment for a year is not implausible.)

A potential second mechanism for slowing a droplet's descent is photophoresis. Photophoresis is the movement of droplets in a light field. Illumination of a particle from one side results in uneven heating, transferring momentum asymmetrically to the surrounding gas and hence moving the particle (Jovanovic, 2009). Negative photophoresis is movement toward the light source. Negative photophoresis has been implicated in keeping aerosol particles aloft in the terrestrial stratosphere (Rohatschek, 1996), and it has been speculated as a mechanism that could increase the time that aerial microorganisms stay aloft in Earth's atmosphere (DasSarma et al., 2020) (see Fig. 1 for a cartoon schematic of this process). In principle, and by analogy with the terrestrial stratosphere, if the inhabited Venusian cloud particles had appropriate optical properties, their fall could be slowed or even prevented by negative photophoresis.

Appropriate properties to generate a negative photophoretic effect would include shape, size, refractive index, and wavelength-dependent absorbance of the particles as well as the spectrum of light impinging on the droplet at different angles. Detailed modeling of this process is beyond the scope of this article, especially as several of the relevant properties of the cloud particles are only assumed from bulk measurements and are not accurately known. Such modeling is for future work, but the existence of the mechanism and its known relevance to terrestrial stratospheric droplets (which can contain high sulfuric acid concentrations (*e.g.*, Bains *et al.*, 2022c) suggests that such modeling would be worthwhile.

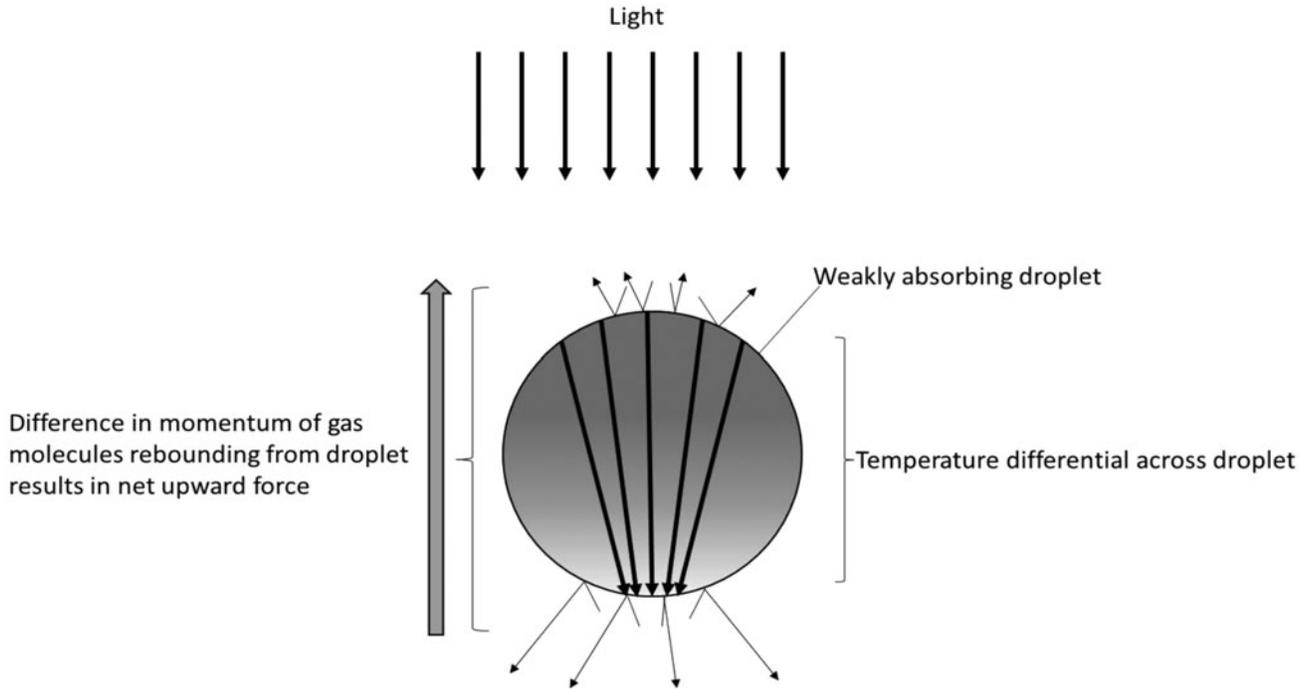

**FIG. 1.** Cartoon schematic of negative photophoresis providing an elevating force on a weakly absorbing droplet under vertical illumination. Negative photophoresis force can keep the life particles aloft.

We note that photophoresis can only occur when there is a light source, and for half of the time the Sun would not be available as a light source. However, the atmosphere of Venus super-rotates (*i.e.*, the atmosphere rotates faster than the planet itself). Zonal windspeeds in the cloud deck have been measured as 40–60 m/s relative to the ground, which means that the clouds are carried around the planet on a timescale of 4–5 days (Read and Lebonnois, 2018). An 8 μm diameter Mode 3 particle would have a settling velocity of $\sim 3 \cdot 10^{-3}$ m/s under Venus cloud conditions (Bains *et al.*, 2021a), and so would fall $\sim 750$ m during the period of darkness. Overall, 99.6% of the cloud particles are smaller than 8 μm, and so will fall more slowly. It is therefore possible that photophoresis could keep cloud particles aloft even if the process only happens half of the time the clouds are carried around the planet, as long as photophoresis can provide a positive lift during the hours of sunlight.

We conclude that terrestrial precedent exists for mechanisms that could keep at least some life-containing cloud particles aloft on Venus, and it prevents the entire ecology from inevitably falling to its doom in the hot, lower layers of the atmosphere.

## 3. Potential Interdictors of Venusian Life Are Insufficient to Rule Out Life

In this section, we address the features of Venus that distinguish its clouds from Earth as a potentially habitable planet. We show that although all of them are significant challenges for Earth life in the clouds of Venus, none are fundamental barriers to the existence of life in the clouds.

### 3.1. Energy

Life requires chemical energy to power its metabolism and physical processes such as movement, division, and transport of molecules into and out of the cell. In this section, we show that there is little direct chemical energy available in the clouds of Venus, but abundant light energy, which could be harnessed, as it is on Earth, to provide the energy that life in the clouds requires.

#### 3.1.1. Energy requirements for life.

Life requires energy for two general classes of biological activity—capture of materials from which to build biomass, and maintenance of that biomass (including movement, feeding, reproduction etc.). The power requirements for biomass capture depend on the rate of growth and the degree to which the metabolisms concerned deploy power for growth. For example, on Earth the same basic metabolism utilizing the same power source for carbon fixation—sunlight—can provide for dramatically different growth rates.

Growing branches of the Pacific yew (*Taxus brevifolia*) increase in diameter by $\sim 0.5\%$/year (Busing *et al.*, 1995), whereas new shoots of the giant bamboo (*Phyllostachys reticulata*) reach full height in less than 2 months, with a maximum growth rate of $\sim 1$ m/day (Li *et al.*, 1998). This is five orders of magnitude difference in growth rate, even though the biochemical basis of growth is the same. This means that growth rate, and hence the rate of growth-driving metabolism in the presence of abundant energy, is constrained primarily by specifics of the ecology of the organism, not by its chemistry. We know that if an organism is energy limited, then that energy limit will limit the maximum rate at which it can grow. However, even in the case of energy limitation, there is no pre-determined minimum growth rate set by energy requirements, as illustrated by the different growth rates of yew and bamboo cited earlier (although there will be other limitations set by other ecological facts, such as predation, chemical degradation, and the rate at which organisms settle out of the cloud layer).

We should also note that the mass of an ecosystem depends on the balance between growth rate and loss rate (*e.g.*, Keeling and Phillips, 2007; Del Grosso *et al.*, 2008). Extremely slow growing, low energy organisms can, nevertheless, accumulate significant biomass if loss rates are small (*e.g.*, like it is in the case of arctic bivalves, which accumulate substantial biomass using very little energy by minimizing biomass loss [Welch *et al.*, 1992]). However, we can calculate the energy needed to maintain a given biomass if we assume that the relationship between maintenance energy and temperature that is widely found in terrestrial microorganisms also applies to Venusian life. This assumption can be made without prejudice as to where that energy comes from (We note that this energy calculation does not include consideration of the energy needed to maintain life in an arid, acid environment. We consider this aspect of maintenance energy separately in Section 3.2. below).

In summary, it has been found (Tijhuis *et al.*, 1993; Hoehler, 2004, 2007; Seager *et al.*, 2013; Hoehler *et al.*, 2020) that the "maintenance energy" (more properly called a maintenance power) $P_{me}$ for microbial growth under a wide range of conditions is given by

$$P_{me} = A \cdot e^{\frac{-E_a}{R \cdot T}}$$

where $P_{me}$ is the energy needed in kJ per gram weight of biomass per second, $E_a = 6.94 \cdot 10^4$ J/mol, $A$ is a constant that varies with metabolism, but is equal to $2.2 \cdot 10^7$ kJ/g of biomass per second for anaerobic metabolism, $R$ is the gas constant = 8.314 J/mol/K, and $T$ is the absolute temperature. See Seager *et al.* (2013) for a more detailed review of the relationship between biomass and maintenance energy.

Microorganisms can maintain themselves in a dormant state with much lower power requirements than the $P_{me}$ calculated above, but a biosphere cannot be composed entirely of dormant organisms. At least some of the biomass must be actively growing to allow the biosphere to survive.

The total power requirements of the various scenarios in Table 1 are summarized in the last column of Table 1. Power does not scale exactly with biomass, as $P_{me}$ is determined by temperature, which varies with altitude, so life that is confined to Mode 3 droplets, which are only found in the lower clouds, requires more energy per gram on average than life that is distributed uniformly in all droplets.

Where could this required energy come from? In principle, energy may be harvested from any gradient—chemical, gravitational, magnetic, thermal etc.—but terrestrial organisms have only evolved mechanisms to harvest chemical and light energy, probably because other sources cannot provide the necessary minimum of energy density within a single cell (Hoehler, 2004, 2007). Life on Venus also is likely to use only light or chemical energy sources, for the same reason.

3.1.2. *Light energy available.* Terrestrial precedent shows that light energy can be captured through several mechanisms, capturing photons in pigments derived from porphyrins (Govindjee and Whitmarsh, 1982; Portis, 1982), retinal (Stoeckenius and Bogomolni, 1982; Béja *et al.*, 2000), melanin (Bryan *et al.*, 2011; Malo and Dadachova, 2019), and carotenoids (Valmalette *et al.*, 2012). Light energy can be used in one of two broad classes of metabolic activity—photosynthesis and phototrophy. Photosynthesis is the use of light energy to drive the reduction of oxidized carbon to biomass. Phototrophy is the capture of light energy to generate adenosine triphosphate (ATP) to drive metabolic reactions.

Blue-green algae and their plant descendants use light for both photosynthesis and phototrophy, not only generating reducing equivalents (and molecular oxygen as a byproduct) but also generating ATP during oxygenic photosynthesis.

Many species of bacteria (*e.g.*, Pierson *et al.*, 1985; Madigan and Ormerod, 1995; Béja *et al.*, 2000; Rappé *et al.*, 2002; Bryant *et al.*, 2007; Zeng *et al.*, 2014; Zervas *et al.*, 2019) as well as some fungi (Gleason *et al.*, 2019), zooplankton (Stoecker *et al.*, 2017), and aphids (Valmalette *et al.*, 2012) are pure photoheterotrophs, using light solely to generate metabolic energy and gathering carbon from other sources. A range of photosynthetic organisms can also switch to a photoheterotrophic metabolism if organic carbon is available in their environment (Béjà and Suzuki, 2008; Stoecker *et al.*, 2017). Thus, on Earth, the use of light solely as a source of energy, and not to power the generation of electrons to reduce $CO_2$, is widespread.

Venus is nearer to the Sun than the Earth, and so intercepts more solar energy; if solar light energy is sufficient to be the primary energy source for life on Earth, it seems logical that it should be able to support a much smaller biosphere on Venus. The flux of visible light (380–740 nm) at the base of the Venusian clouds (47 km altitude) is $\sim 63.6$ J/m$^2$/sec, or $7.32 \cdot 10^{15}$ J/planet/second. Even if the clouds were composed of 6.8% biological matter (the highest loading of the models summarized in Table 1), the maintenance energy required for that biomass would only require the capture of 0.01% of the incident sunlight. As noted earlier, this source of energy would only be available half of the time the clouds are carried around the planet, so organisms would have to capture twice as much energy during illumination and store energy for subsequent deployment during darkness, as phototrophs do on Earth.

3.1.3. *Chemical energy available.* By contrast, chemical energy is quite limited in Venus' consensus atmosphere compared with the light energy available (Cockell *et al.*, 2021a; Jordan *et al.*, 2022). Jordan *et al.* (2022) focused on whether the $SO_2$ depletion observed through the clouds could be directly explained by life using any of three specific sulfur-based energy metabolisms as their sole source of energy. They tested the potential for three kinds of metabolism initially suggested by Schulze-Makuch and Irwin (2006) and Schulze-Makuch *et al.* (2004) to support a biosphere and to explain the depletion of $SO_2$ observed in the cloud decks.

All the metabolisms required reduced species as input, either hydrogen-containing compounds ($H_2S$ or $H_2$) or carbon monoxide. Reduced compounds are predicted to be rare in Venus' oxidized atmosphere (*e.g.*, Marcq *et al.*, 2018), and the limited experimental data confirm this (*e.g.*, Mogul *et al.*, 2021b; 2022), so it is not surprising that the chosen metabolisms could not explain the $SO_2$ depletion. Interestingly, Jordan *et al.* found that the chemical energy available could support a biosphere within the mass ranges summarized in Table 1.

Jordan *et al.* (2022) did not consider other sulfur-based energy metabolisms that could both explain the $SO_2$ depletion and explain the presence of sulfur-rich ''haze'' that likely extends from altitudes above the clouds to sub-cloud layers. For example, heterotrophic oxidation of biomass by sulfur dioxide, which might be characterized as

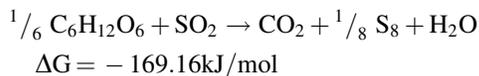

$$^1/_6 \, C_6H_{12}O_6 + SO_2 \rightarrow CO_2 + {}^1/_8 \, S_8 + H_2O$$
$$\Delta G = -169.16 \, kJ/mol$$

(here illustrated as the oxidation of glucose: $\Delta G =$ standard free energy of reaction in aqueous solution at 298 K—data from Amend and Shock [2001]) removes $SO_2$, generates sulfur, and releases substantial energy. Thus, heterotrophic oxidation of biomass (predation) might explain the removal of $SO_2$ and the generation of $S_8$ (We note that a biosphere cannot be explained entirely by heterotrophic life; there must be primary producers that make the biomass for heterotrophs to eat).

Other mechanisms have also been suggested to solve both the $SO_2$ depletion ''problem'' and other unexplained observations in Venus' atmosphere (Bains *et al.*, 2021a; Rimmer *et al.*, 2021). Thus, although Jordan *et al.*'s (2022) conclusions that three specific sulfur-based energy-metabolisms cannot explain the atmospheric chemistry of Venus are correct, they cannot be extended beyond those modeled metabolisms to any general statement about Venus.

We conclude that there is abundant energy for life in the clouds of Venus.

### 3.2. Low water activity of the clouds

Venus is usually understood to be extremely dry, unlike any terrestrial environment. The atmosphere is dry in the sense of having very little chemically available water. The extremely low water activity has been suggested as an insuperable barrier to the presence of life, which requires abundant available water. In this section we argue that this is, indeed, a major barrier to our understanding of how life could operate in the clouds of Venus but cannot *a priori* rule out the presence of life.

There is substantial disagreement concerning the water content of the atmosphere from modeling and from measurements (reviewed in De Bergh *et al.* [2006]), but all authorities agree that the atmosphere is extremely dry. One reason for considering that the clouds are composed of concentrated sulfuric acid, rather than dilute acid, is the very low water activity in the atmosphere. Sulfuric acid is extremely hygroscopic; only in an extremely dry environment would concentrated acid be stable to absorption of water and consequent dilution.

The cloud particles are modeled to contain up to 20% water by weight, but this water is tightly bound to sulfuric acid and is not available to act as a solvent. Therefore, it is more accurate to state that the Venusian atmosphere, and the cloud droplets that are presumed to be in equilibrium with that atmosphere, have very low water activity. The very low water activity absolutely rules out the possibility that any known terrestrial life would flourish on Venus. Terrestrial life is generally considered not to be able to grow at water activities ($a_w$) below ∼0.58 (Fontana, 2020), whereas the average Venusian water activity is nearer 0.002 (Hallsworth *et al.*, 2021). Several authors have used this as an argument for the uninhabitability of Venus (Cockell *et al.*, 2021a; Hallsworth *et al.*, 2021). This, however, misses two points.

First, the arguments are based on adaptations of known terrestrial life, where water is almost universally abundant, even if only transiently. Environments where $a_w$ falls below 0.6 are widespread, and in principle any organism adapting to be able to grow at low $a_w$ would have a selective advantage in environments such as the Atacama or Sahara deserts. So why is there no terrestrial life that can flourish in these extensive environments, rather than just surviving there in a dormant state and growing in rare wetting events? The answer may lie in the word ''transiently.''

Adaptation of growth in low $a_w$ requires adaptation of every aspect of biology in which water has a role, which is every cellular process. Such adaptation is a specialization, and it comes with consequent reduced fitness when growing in high $a_w$ environments. There are abundant examples of xerophilic organisms that can grow at low $a_w$ but grow poorly at higher $a_w$ (*e.g.*, Pitt and Hocking, 1977; Su-lin *et al.*, 2011; Stevenson *et al.*, 2017; reviewed in Brown [1976]).

A cornucopia of examples exists of organisms forming dormant forms to survive without growth in hostile environments. However, the formation of all these dormant forms of life require time to switch from active growth to dormancy. As a well-studied example, spore-formation in *Bacillus subtilis* in response to nutrient deprivation takes 1–5 days (Armstrong *et al.*, 1970). If lethal environmental change happens very fast, such adaptive changes cannot be made, and the organism dies. On Earth, rainfall (however infrequent) is effectively instantaneous; one minute an organism is in a dry environment, the next in a wet one. If the organism is so highly adapted to a low water activity environment that it cannot function in a wet environment, then rainfall will kill it.

This phenomenon is, indeed, observed in some highly xerophilic Atacama microorganisms (Azua-Bustos *et al.*, 2018). It is, therefore, the unpredictable changes in the environment that predominantly set the limits of life, or make the environment uninhabitable, rather than the average absolute values of environmental parameters. The lower limit of $a_w = 0.58$ for terrestrial organisms is, therefore, not a reflection that terrestrial biochemistry adapted to lower $a_w$ is inconceivable, but that life cannot develop a chemistry that can function at low $a_w$ and can also function at high $a_w$. As drying out after rainfall takes time, the adaptive solution is therefore to remain dormant at low $a_w$ and wait for rain.

Venusian cloud environment is different. It is permanently extremely arid, an environment for which there is no terrestrial precedent. Were life to exist there, it would not have to adapt to survive sudden periods of wetting and high $a_w$. Evolutionary selection pressures would be fundamentally different from those on terrestrial xerophiles. Evolution of an organism that could grow at $a_w = 0.01$ but is killed by $a_w = 0.1$ would be highly favored on Venus. Thus, the argument that life on Earth is not known to grow at $a_w < 0.58$ cannot be used to argue that life on Venus cannot grow at $a_w < 0.58$, and hence that there cannot be life in the clouds of Venus.

Second, the water activities cited for Venus' atmosphere are averages, often integrated over large altitude ranges and made without knowledge of regional gradients, not actual local measurements. Figure 2 illustrates the range of measured water abundances and the range of model predictions for water abundance in Venus' atmosphere. Specifically, *in situ* measurements have suggested much higher water activities in some regions of the clouds (Petkowski *et al.*, 2023), which suggest the presence of relative ''wet zones'' in the atmosphere. Although these ''wet zones'' are extremely dry by comparison to any environment on the surface of the Earth, they may provide a less hostile environment than the average water activity of the atmosphere suggests.

Could an active biosphere exist, in principle, in an environment with very low water activity, independent from the undoubted fact that no known terrestrial life could flourish there? We can start to address this challenge as follows. We assume life is cellular, with an aqueous interior. It must therefore be surrounded by a wall or membrane that is relatively impermeable to water but sufficiently permeable to other materials, by either passive diffusion or active transport. It is implausible that such a membrane would be completely impermeable to water, so the cell would have to expend energy in pumping water from an exterior where water activity was very low into an aqueous interior at a rate that balances the leakage of water from the cell interior to the exterior environment. Is such pumping possible?

We can estimate the rate at which a cell could pump water into the cell interior from the available energy as follows. We assume the cell is a sphere, so the surface through which water can diffuse is four times its cross-sectional area regardless of the cell's size. We assume the source of energy is sunlight, and therefore the amount of energy captured is proportional to cross-sectional area of the cell.

The energy required to move water from the external atmosphere is given by

$$\Delta G = R \cdot T \cdot \ln \frac{\{H_2O\}_i}{\{H_2O\}_e} \quad (1)$$

where $\Delta G$ is the energy in joules, $R$ is the gas constant = 8.314 J/K, $T$ is the absolute temperature, $\{H_2O\}_i$ is the internal activity of water, which we assume = 1, and $\{H_2O\}_e$ is the external activity of water. If we assume that the droplets and the atmosphere that surrounds them are in chemical equilibrium, then the water activity outside a cell suspended in a droplet is the same as the water activity in the atmosphere around that droplet, and so $\{H_2O\}_e$ can be approximated by

$$\{H_2O\}_e = \frac{pp(H_2O) \cdot P}{VP_{H_2O}} \quad (2)$$

where $pp(H_2O)$ is the partial pressure of $H_2O$ in Venus' atmosphere, $P$ is the atmospheric pressure, and $VP_{H_2O}$ is the vapor pressure of water over pure liquid water at that temperature. The vapor pressure of water over pure liquid water was calculated according to the Antoinine equation provided by (Yaws, 1999) and summarized in Eq. (3)

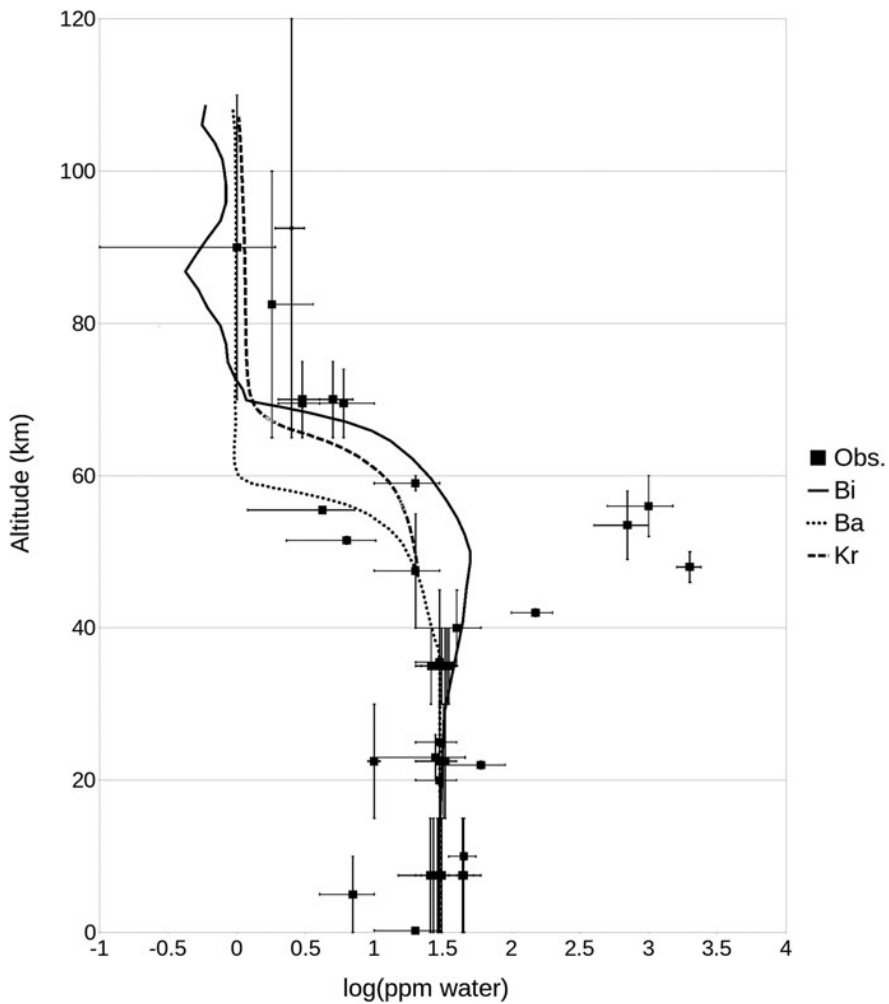

**FIG. 2.** There is a substantial range of measured and modeled water abundance in the gas phase of Venus' atmosphere. X axis: $\log_{10}$ of ppm water. Y axis: altitude in km. Kr = Krasnopolsky (2012) photochemical model of the atmosphere above 47 km. Bi = Bierson and Zhang (2019) photochemical model. Ba = Bains *et al.* (2021a) photochemical model assuming ammonia production to neutralize acid. Obs = observations of water abundance from Table 1 of Rimmer *et al.* (2021). Reported abundances are often reported as the average value over a range of altitudes where the measurements were made (under the assumption that water abundance does not change significantly over that range); error bars are for the range of latitudes to which that measurement pertains (vertical error bars), and the reported range of water abundance (horizontal error bars). In the cloud level (48–60 km altitude) modeled and measured water abundances vary by four orders of magnitude.

$$log_{10}(VP_{H_2O}) = A + \frac{B}{T} + C \cdot log(T) + D \cdot T + E \cdot T^2 \quad (3)$$

where A through E are constants such that $A = 29.8605$, $B = -3152$, $C = -7.3037$, $D = 2.42 \cdot 10^{-9}$, $E = 1.81 \cdot 10^{-6}$, T is the absolute temperature, and $VP_{H_2O}$ is calculated in millimeters of mercury. Atmospheric pressure and temperature as a function of altitude was taken from Venus International Reference Atmosphere (VIRA), and light energy square meter between 380 and 740 nm was kindly provided as an output of the radiative transfer model described in (Rimmer et al., 2021; Jordan et al., 2022).

We know from the calculation cited earlier that only a very small fraction of incident sunlight (0.01%) is needed to provide the maintenance energy $P_{me}$ required for continued vitality, so most of the energy available from sunlight could be used for re-capturing water. If we assume 10% of the energy in sunlight between 380 and 740 nm is used for water pumping, we can calculate the rate at which a spherical cell could pump water, by Eq. (4)

$$Flux = \frac{L}{\left(R \cdot T \cdot ln \frac{VP_{H_2O}}{pp(H_2O) \cdot P}\right).4} \quad (4)$$

where L is the light energy per meter squared per second, and the factor 4 is because the surface of a spherical cell has four times the cross-sectional area, so there is four times as much area to pump water across as there is area to capture light. This is illustrated in Fig. 3.

The pumping rates shown in Fig. 3 are quite low rates of water loss. This is the rate of loss that can be balanced by pumping water back into the cell, but for reference, if water was pumped into a bacterial spore of volume $\sim 0.5$ fL (Nicholson et al., 2000) at 0.001 moles/m²/second, without loss back to the outside, enough would be pumped to fill a living bacterial cell of volume 4.6 fL (Yu et al., 2014) in around 500 s. Such low rates of water loss are not unfeasible to imagine.

Human skin, which is not optimized for water retention, loses around 0.00035 moles/m²/second of water into air as dry as Venus' through "insensible perspiration" (Buettner, 1953), and desert ants lose around $2.6 \cdot 10^{-5}$ moles/m²/second through their cuticle (Lighton and Feener, 1989). Both skin and cuticle are complex, multi-layered structures, but it is not unphysical to suggest that nanometer-thick boundary materials (not necessarily Earth-like lipid bilayers) could have a lower permeability, and we illustrate this with the example of graphene.

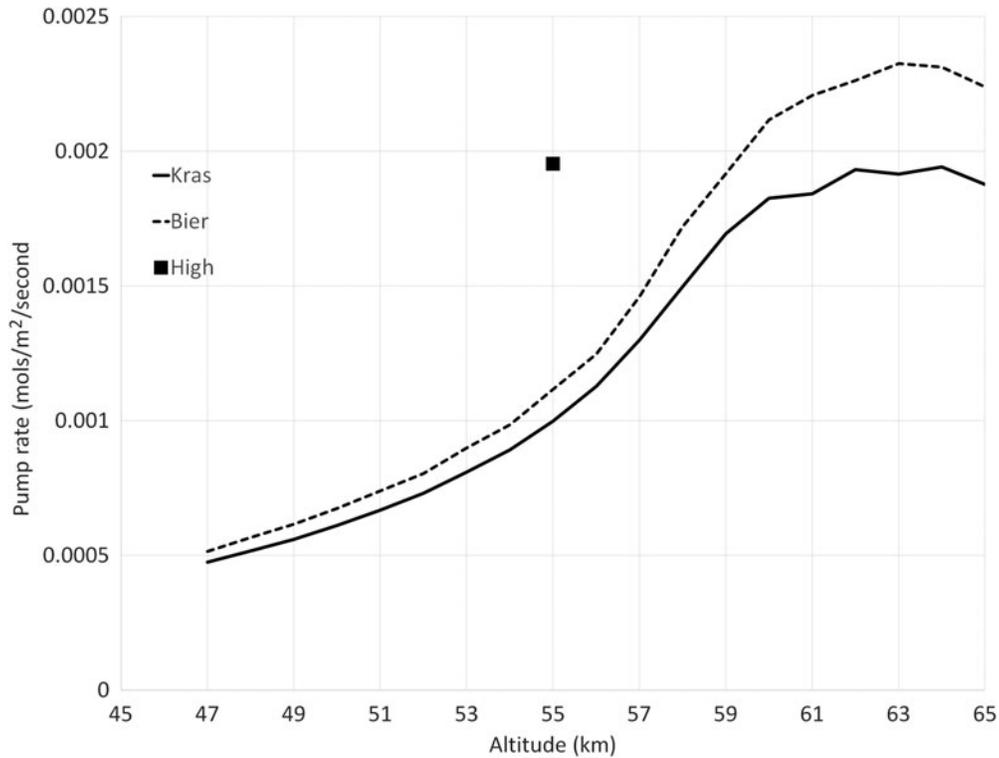

**FIG. 3.** The maximum rate at which a cell could pump water into itself from the dry Venusian atmosphere is substantially greater than the leak rate through single layer graphene. X axis: altitude (km). Y axis: Rate at which water could be pumped against a concentration gradient if 10% of the ambient light energy was used. Kras = assuming the water abundance of Krasnopolsky (2012). Bier = assuming the water abundance of Bierson and Zhang (2019). The black square point ''High'' is assuming a 1000 ppm water value at 55 km, as suggested by some of the observationally determined values in Fig. 2. Details of the calculations presented in Fig. 3 are available in the Supplementary Data S1.

Single-layer graphene has a permeability to water of $\sim 10^{-6}$ g/m²/day (Seethamraju et al., 2016; Kwak et al., 2018), greatly below any of the values in Fig. 3. We do not wish to postulate organisms entirely encased in graphene; apart from any other consideration, graphene shows poor stability to acid (Yu et al., 2016). However, this extreme example illustrates that an organism that can harness light energy to counteract the leakage of water from an aqueous interior into Venus' atmosphere appears not to be physically impossible.

We conclude that the aridity of the atmosphere is a major barrier to life, and in our view the most important one as energy is clearly abundantly available. However low water activity cannot *a priori* rule out the possibility of life. This argument is distinct from the more general argument that Venus' atmosphere lacks hydrogen atoms, which we address next.

### 3.3. Lack of hydrogen

All biochemistry requires an abundance of hydrogen atoms, quite apart from its need for water. However, Venus' atmosphere is very short of compounds that contain hydrogen atoms. The lack of hydrogen has been postulated as a reason that biochemistry is unlikely in Venus' atmosphere (Benner, 2021). We next show that shortage of hydrogen only imposes a moderate energy cost on life in the clouds, and it is not a substantial limiting factor.

The lack of water in the atmosphere of Venus reflects a lack of hydrogen atoms in the planet's atmosphere and crust. Hydrogen atoms are central to terrestrial biochemistry, where not only do they comprise $\sim 50\%$ of the atoms in terrestrial biochemicals (Bains et al., 2021a), but they are also key players in terrestrial biochemistry through their role in chemical functionalities such as amines, aldehydes, carboxylic acids etc., in forming hydrogen bonds, and as mediators of energy metabolism through proton gradients.

It seems reasonable to argue that any life must have an abundant source of hydrogen atoms to build biochemistry, regardless of the solvent it uses. The components of Venus' atmosphere contain very little hydrogen: the sum of all the H atoms in the atmospheric gases amounts to $0.08 - 0.24\%$ of the total atoms (depending on the gas abundances, as described in Bains et al. [2021b]). The low abundance of hydrogen-containing atmospheric gases is a barrier to life in two ways. First, it means that water is very scarce, a specific barrier that we discussed earlier (Section 3.2.). Second, and more generally, it means that life must expend energy retrieving hydrogen from its environment, rather than (as on Earth) regarding hydrogen as an abundant element that can be had "for free." However, this means that the lack of hydrogen is a quantitative, and hence an energetic, impediment to life, not an absolute barrier, as we now illustrate.

Table 2 shows the overall energetics of chemical transformations that capture hydrogen into reduced molecules; we note that these are not metabolic pathways, only the summary input and output of metabolic pathways. The actual energy for performing these transformations are likely to be higher for a real metabolism. The capture of hydrogen from water as reduced carbon, sulfur, or nitrogen has an energetic cost, but it is small compared with the cost of reducing oxidized elements.

Specifically, the cost of reducing $CO_2$ to glucose is 6.5% higher on Venus than on Earth if we consider just the input and output of the net chemical transformation and assume the abundances of other reactants and products are the same between Earth and Venus. The increased energetic cost of capturing carbon as glucose is due to the energetic cost of capturing rare water, compared with the energetic cost of reducing $CO_2$ in an environment where water is abundant (We note that the ceteris paribus assumption is not true; for example, $CO_2$ is much more abundant on Venus, making any $CO_2$-consuming reaction more thermodynamically favorable than on Earth, so the calculations presented in Table 2 represent a conservative approach).

This result might at the first glance appear to be unexpected. The relatively small amount of energy needed to capture hydrogen from an environment in which hydrogen is rare is the result of the form of Eq. (1) cited earlier. The energy needed to capture any substance against a concentration gradient is a logarithmic function of the concentration difference. As a result, capturing hydrogen against a concentration difference of 1:10,000 at 298 K requires 22.8 kJ/mol. By contrast, reducing the extremely stable carbon-oxygen to a carbon-hydrogen and a carbon-alcohol group in glucose requires ∼220 kJ/mol.

Table 2. Comparative Energetics of Hydrogen Capture Reactions on Earth and Venus[a]

| | | Free energy of reaction (kJ/mol) | |
|---|---|---|---|
| Chemical process | Reaction | Assuming Earth $H_2O$ abundance | Assuming Venus $H_2O$ abundance |
| Carbon fixation | $CO_{2(aq)} + H_2O_{(g)} \rightarrow \tfrac{1}{6} C_6H_{12}O_{6(aq)} + O_{2(g)}$ | 424.0 | 451.9 |
| Sulfur fixation | $SO_{2(aq)} + H_2O_{(g)} \rightarrow H_2S_{(aq)} + 1\tfrac{1}{2} O_{2(g)}$ | 434.1 | 462.0 |
| Nitrogen fixation | $\tfrac{1}{2} N_{2(g)} + 1\tfrac{1}{2} H_2O_{(g)} \rightarrow NH_{3(aq)} + \tfrac{3}{4} O_{2(g)}$ | 257.8 | 285.7 |

[a]Note that the actual biochemistry of sulfur and nitrogen fixation in terrestrial biochemistry is a combination of photosynthesis and subsequent use of reducing power to reduce sulfur or nitrogen respectively. All reagents are assumed to have abundances seen on Venus and as described in (Bains et al., 2021b) except water, which has activity = $1.3 \cdot 10^{-5}$ and is assumed to be in gas phase on Venus, while on Earth we assume water activity = 1. Ammonia was assumed to be present at 1 ppm; glucose at 1 mM. Calculations were performed for temperature = 298 K, pressure = 0.5 atm., which approximate the conditions at 55 km altitude on Venus, in the middle of the cloud layer. Thermodynamic data from (Amend and Shock, 2001).

We conclude that the overwhelming energetic cost of building biomass is not finding and capturing hydrogen, but reducing carbon, assuming life uses $CO_2$ as a carbon source. The rarity of hydrogen atoms does not render the Venusian clouds uninhabitable.

### 3.4. Acidity of the clouds

The clouds of Venus have been hypothesized to be made of concentrated sulfuric acid, a model that has been accepted as canonical for 50 years despite a range of anomalies (Bains *et al.*, 2021a; Mogul *et al.*, 2021a; Petkowski *et al.*, 2023). However, the presence of liquid concentrated sulfuric acid is an inference, not a measured fact, and a number of measurements are inconsistent with the clouds being solely composed of concentrated sulfuric acid droplets (Bains *et al.*, 2021a; Petkowski *et al.*, 2023).

If the clouds are mostly made of concentrated sulfuric acid, then life could, in principle, exist in them in one of three ways. First, life could use energy to abstract water from the sulfuric acid and maintain an internal milieu that is water-based. This would require a cell boundary (membrane or wall) that was impermeable to sulfuric acid, resistant to acid on its outer face, and yet permeable to other nutrients. Terrestrial proteo-lipid membranes do not approach these specifications.

Survival in concentrated sulfuric acid would also require substantial energy expenditure to keep water inside the cell against leakage, although we have argued earlier (Section 3.2.) that more than sufficient energy is available to provide for water pumping. It is possible that the cells could have a multi-layered wall structure around them to provide additional defense against acid, possibly a wall that was sacrificially cross-linked by the acid itself to an acid-resistant polymer. An example of such a solution could be layers of acid resistant membranes composed of Earth-lipid analogue compounds resistant to concentrated sulfuric acid attack (see [Seager *et al.*, 2021a], their Appendix A, for preliminary results on the formation of such structures).

Second, life could neutralize the acid. Bains *et al.* (2021a) have provided a model of how this could work, cite precedent in terrestrial biology, and provide arguments for why the Mode 3 particles in the clouds may actually be composed of partially neutralized salts of sulfuric acid, not the liquid sulfuric acid itself. We note that the neutralization of acid does not address the related challenge of the extreme aridity of the clouds (see Section 3.2.).

Lastly, and very speculatively, life could use concentrated sulfuric acid itself as a solvent. Specifically, a structurally and functionally diverse set of small organic molecules and macromolecules are predicted to be stable in concentrated sulfuric acid under Venusian cloud conditions (Bains *et al.*, 2021c). Concentrated sulfuric acid can also support insoluble polymers and amphiphiles that could potentially form membrane structures (Bains *et al.*, 2021c). Although life based on a solvent other than water, and specifically on concentrated sulfuric acid, is in our view highly speculative, it cannot be ruled out *a priori*.

We conclude that the acidity of the clouds may be a substantial obstacle for life to overcome, but at least two routes to overcome it can be imagined by analogy with terrestrial life: forming an acid-resistant outer wall or membrane and neutralizing the acid. If life neutralizes the acid in droplets, it would both explain several anomalies in Venus' atmosphere (Bains *et al.*, 2021a) and obviate any other solution.

### 3.5. Lack of metals

Life on Earth is obligatorily dependent on metals for catalysis, electron transfer, and molecular structure (Hoehler *et al.*, 2020). Life on Earth derives metallic elements from the crust, either by direct biological weathering or through abiotic solution of minerals into surface water. Even aerial microbial life on Earth relies on many metalloenzymes and has systems for metal capture (Amato *et al.*, 2019). On Venus, an aerial biosphere will not have access to either the surface or surface water (as there is no surface water). How then could it access metals?

It is quite likely that the lower atmosphere of Venus contains volatile metal compounds, specifically iron chloride, which could be boiled off the surface and condense at cloud altitudes into aerosols (Krasnopolsky, 2017). Iron species have been detected in the clouds of Venus (Petrianov *et al.*, 1981). Some alkaline metals, some transition metals apart from iron, and many metalloids such as selenium also have halides or oxides that are volatile at Venus surface temperatures (Marov and Grinspoon, 1998). This includes molybdenum (V) chloride, with a boiling point of 268°C (Speight, 2017). Molybdenum is required by terrestrial life for nitrogen reduction (Presta *et al.*, 2015), a reaction that is a key component of the model that suggests that Venusian life neutralizes cloud acid with ammonia (Bains *et al.*, 2021a).

Delivery of other metals, which do not form volatile compounds under Venusian surface temperature conditions, to the clouds would rely on explosive volcanism (which is likely to be rare on Venus [Bains *et al.*, 2022b]), transport of dust from the surface to the clouds (Sagan, 1975; Rimmer *et al.*, 2021), or meteoritic infall (Bains *et al.*, 2021b; Omran *et al.*, 2021). Both could deliver small amounts of all the metallic elements to the clouds.

Whether the rate of delivery by volcanism, dust, or meteoritic infall is sufficient to sustain an aerial biosphere depends on how efficiently the biosphere can retain metals. Organism corpses lost to settling will leave their metallic content on the surface. Elements that cannot be volatized will then be irreversibly lost to the surface until they are eventually recycled by volcanism, which is likely to be extremely inefficient at returning surface elements to the clouds (Bains *et al.*, 2022b). Thus, for non-volatizable elements, Venusian organisms will be limited by re-delivery from dust or meteoritic infall. Whether this is sufficient will depend on how much of each non-volatile metal each cell needs, which is not knowable *a priori*.

We also note that it is not clear what metals are essential for life. It is likely that several redox-active elements such as iron and molybdenum, and several ''hard'' non-redox active metals such as sodium and magnesium to act as ligands and charge carriers are universally essential to allow the complex chemistry of life (Da Silva and Williams, 2001). However, terrestrial life shows that there is surprising flexibility in swapping between metals with similar functions within terrestrial life (Hoehler *et al.*, 2020). Thus, the lack of a volatile source, for example, cobalt or nickel, need not be a show-stopper for a Venusian biochemistry.

We conclude that the availability of metals to Venusian cloud life poses different problems to those faced by terrestrial organisms (Li *et al.*, 2020), some of which are unknown as the metal chemistry of the clouds is poorly constrained, but the delivery of minerals from dust and meteorites combined with the likely flexibility of biochemistry to use the metallic resources available suggests that mineral availability is not likely to be a limit on the presence of cloud-based life on Venus, although it may limit its abundance.

### 3.6. High radiation environment

Venus is closer to the Sun than the Earth, and the clouds are high in the atmosphere, both of which might suggest that ultraviolet (UV) radiation, and potentially solar X-rays or cosmic rays, might render the clouds uninhabitable, as they render the surface of Mars uninhabitable to life as we know it (Dartnell *et al.*, 2007). However, initial radiative transfer modeling of the clouds suggests that enough UV is absorbed to make the clouds below 59 km potentially habitable, and an optimal balance between incidence visible light (for photosynthesis) and reduced UV and ionizing radiation is at 54 km (Dartnell *et al.*, 2015; Patel *et al.*, 2021).

These models include the "unknown UV absorber" (Titov *et al.*, 2018), which predominantly absorbs UV light in the top clouds at ~60 km and has been speculated to be a product of life (Limaye *et al.*, 2018). Although the exact altitude considered potentially habitable is likely to be revised by more detailed modeling, for example, modeling that includes variations in altitudinal abundances of UV absorbing species considered constant in the original model (*e.g.*, Sandor and Clancy, 2017, 2018), the overall conclusion—that there is a mid-cloud altitude below which life could be protected from radiation damage—is likely to remain. This conclusion has also recently been supported by radiation dosimetry calculations for the Venusian atmosphere during different periods of solar activity (Tezari *et al.*, 2022). We conclude that the Venusian cloud radiation environment is not particularly hostile to life.

## 4. Why Consider Venusian Life?

In section 3, we have argued that none of the objections to the concept of life in the clouds of Venus stands up to quantitative scrutiny, unless we make the unwarranted assumption that Venusian life must be essentially the same as terrestrial life.

The community considers that it is unlikely that there is life on Venus (Bains and Petkowski, 2021). So why consider such a possibility at all, when such research might distract attention and resources from searches on targets that the scientific community agrees are more promising, such as Mars or Europa? We believe there are two compelling reasons for considering, modeling, and searching for life on Venus.

First, if the clouds of Venus do host indigenous life, it is likely to be substantially chemically different from Earth life. Venusian life would provide a critical test of what aspects of terrestrial biochemistry could be "universal" and what are contingent on Earth's environment and history. Such comparisons are central for the search for life on other worlds. This is true whether life originated independently on Venus or whether it had a common origin with terrestrial life but has diverged over geological timescales to adapt to the very different environment of Venus. Finding such life would therefore be of pivotal importance in understanding the possible nature and extent of life elsewhere in the Universe. Even a small possibility of making such a discovery is worth an effort to explore. Equally, finding that Venus is indeed uninhabited would constrain our search for life elsewhere.

Second, considering what Venusian life might be, where it might be, and how one would detect it would allow observations of Venus (remote or *in situ*) to be designed to include the possibility of life detection alongside other mission goals. At minimum, future observations should be capable of detecting chemical anomalies, which are likely to be the first indication of the presence of life (Cleland, 2019). Venus' atmosphere is already known to display several unexplained observations (Bains *et al.*, 2021a; Cleland and Rimmer, 2022; Petkowski *et al.*, 2023), such as the chemical nature of the UV absorber (Titov *et al.*, 2018) and the possible presence of highly reducing gases (Greaves *et al.*, 2021b; Mogul *et al.*, 2021b), so any additional information about the atmosphere that explores its chemistry as widely as possible would be of general value.

Such missions could be relatively small in scale. For example, the Rocket Lab mission is being prepared to examine the physical properties of the clouds of Venus (French *et al.*, 2022). The Rocket Lab mission scientific payload is an autofluorescence nephelometer that contains a fluorescence detector to probe whether cloud particles fluoresce in the wavelengths typical of organic matter (Baumgardner *et al.*, 2022). Such fluorescence would be an indication of another chemical anomaly in the cloud chemistry of Venus. Such missions can be developed fast and launched economically to answer specific, focused science questions. The specific, focused answers provided by such missions will not, of course, resolve all the many questions about the complex clouds of Venus, but they will complement and inform larger missions with broader remit and greater analytical power that have longer timescales (NRC, 2012).

We will not know if organic compounds are the products of life or are created abiotically. However, their detection would prove that complex organic chemistry can stably exist in the cloud droplets of Venus' clouds. Such a discovery would change the paradigm that the clouds are inherently incompatible with any complex chemistry and support the idea that the Venus cloud droplets—incredibly harsh for any Earth life—could be habitable to life based on a different biochemistry. But if we do not look for complex organic chemistry, then we will certainly not find it.

## 5. Conclusions

Venus is a planet that is only Earth's sister in mass and overall composition. In terms of habitability, it is more Earth's "ugly step-sister," with an acid cloud deck with extremely low water activity as the only potentially habitable environment. Prospects for life there seem poor, and several authors have concluded that life there is impossible in principle. This article lays out the reasons that we disagree with that assessment. In terms of the availability of energy, Venus' aridity, low hydrogen abundance, and acidity, and the availability of metals, Venus is an extreme environment, and one in which no known Earth organisms with Earth-like biochemistry could survive. But we do not find anything in Venus' environment that precludes life, based on the principles of what we know of life on Earth. Terrestrial life cannot survive in Venus' clouds. We encourage others to consider Venus as a place where some highly non-terrestrial life just might live, and to explore what that life might be and how we might economically search for it.

## Acknowledgment

The authors would like to thank the anonymous reviewers for valuable comments and suggestions.

## Authors' Contributions

Conceptualization: W.B., J.J.P., and S.S.; methodology: W.B.; analysis W.B., J.J.P.; writing—original draft preparation, W.B.; writing—review and editing, W.B., J.J.P., and S.S. All authors have read and agreed to the published version of the article.

## Author Disclosure Statement

No competing financial interests exist.

## Funding Information

This research was partially supported by Breakthrough Initiatives and the Change Happens Foundation.

## Supplementary Material

Supplementary Data S1

Address correspondence to:
*William Bains*
*Department of Earth, Atmospheric*
*and Planetary Sciences*
*Massachusetts Institute of Technology*
*77 Massachusetts Avenue*
*Cambridge, MA 02139*
*USA*

*E-mail:* bains@mit.edu

**Abbreviations Used**

ATP = adenosine triphosphate
N/A = not applicable
UV = ultraviolet
VIRA = Venus International Reference Atmosphere